\documentclass[aps,twocolumn,showpacs,preprintnumbers,floats]{revtex4}

\usepackage{graphicx}
\usepackage{dcolumn}
\usepackage{bm}
\usepackage{epsfig}

\begin{document}

\title{
 Thermodynamics with density and temperature dependent particle masses
 and properties of bulk strange quark matter and strangelets}
\author{
X.\ J.\ Wen,$^1$ X.\ H.\ Zhong,$^2$ G.\ X.\ Peng,$^{\mathrm{1,4}}$
 P. N. Shen,$^{3,1}$
 P. Z. Ning$^{3,2}$
       }
\affiliation{
$^1$Institute of High Energy Physics,
   Chinese Academy of Sciences, Beijing 100049, China   \\
$^2$Department of Physics, Nankai University, Tianjin 300071, China\\
$^3$China Center of Advanced Science $\&$\ Technology (World Lab.),
   Beijing 100080, China  \\
$^4$Center for Theoretical Physics,
 Laboratory for Nuclear Science and Department of Physics,
 Massachusetts Institute of Technology,
 77 Mass.\ Ave., Cambridge, MA 02139-4307, USA
              }

\begin{abstract}
Thermodynamic formulas for investigating systems with density and/or
temperature dependent particle masses are generally derived from the
fundamental derivation equality of thermodynamics.
Various problems in the previous treatments are discussed and modified.
Properties of strange quark matter in bulk and strangelets at both zero
and finite temperature are then calculated based on the new thermodynamic
formulas with a new quark mass scaling, which indicates that low mass
strangelets near $\beta$\ equilibrium are multi-quark states with an
anti-strange quark, such as the pentaquark ($u^2d^2\bar{s}$)
for baryon nmber 1
and the octaquark ($u^4d^3\bar{s}$) for dibaryon etc.
\end{abstract}
\pacs{24.85.+p, 12.38.Mh, 12.39.Ba, 25.75.-q}

\maketitle

\section{Introduction}

One of the most exciting possibility from QCD is that
hadronic matter undergoes a rich and varied phase landscape
with increasing densities.
At extremely high densities, the mass of strange quarks becomes
unimportant and all the three flavors of $u$, $d$, and $s$ quarks
can be treated on an equal footing. Consequently, quark matter
is in the color-flavor-locked (CFL) phase \cite{Rajagopal2001PRL86}
and/or a new gapless CFL phase (gCFL) \cite{Alford2004PRL92}.
However, if densities are not that high, the strong interactions
between quarks become important, and quark matter is in the
unpaired phase. Presently, RHIC is teaching us about the properties
of the hot but not asymptotically hot quark gluon plasma
 \cite{Gyulassy2004,Shuryak2004}.
Actually the future FAIR project which will be built in
the coming years at GSI in Germany is targeted towards the
physics of ultradense matter like it is found in neutron stars
\cite{FAIRweb}.

The original idea of Witten is that strange quark matter (SQM),
rather than the normal nuclear matter, might be the true ground
state of the strong interaction \cite{Witten1984PRD30}.
Immediately after Witten's conjecture, Farhi and Jaffe showed that
SQM is absolutely stable near the normal nuclear saturation density
for a wide range of parameters \cite{Farhi1984PRD30}.
Now SQM has been investigated for
more than two decades since the pioneer works of many authors
\cite{Bodmer1971PRD4,Chin1979PRL43,Witten1984PRD30,Farhi1984PRD30}.
Because the lattice gauge theory still has difficulty in the
consistent implementation of chemical potential in numerical
simulations presently \cite{Creutz2001NPB94c} while the
perturbative approach is unreliable at the strong coupling regime,
phenomenological models reflecting the characteristics of
QCD are widely used in the study of hadrons, and many of them
have been successfully applied to investigating the stability
and properties of SQM.

An important question in the study is how to incorporate
pressure balance. Basically, one uses a model to give the
thermodynamic potential, then add to it a constant
to get mechanical equilibrium. This is the famous `bag' mechanism.
Many important investigations have been done in this direction
\cite{Berger1987PRC35,Gilson1993PRL71,Madsen1993PRL70,%
Madsen2000PRL85,SchaffnerBielich1997PRD55,Parija1993PRC48,%
HeYB53PRC1903,Madsen87PRL172003,Schafer2002NPA,Ratti2003,%
Wangp2003PRC67,Harris2004JPG30}.
A common feature of these investigations is that quark masses are
constant, so the normal thermodynamic formulas can be used without
thermodynamic inconsistency problems.
Actually, however, it is well known in nuclear physics that particle
masses vary with environment, i.e., the density and temperature.
Such masses are usually called effective masses
\cite{Walecka1995OSNP16,Henley1990NPA617,Brown1991PRL66,Cohen1991PRL67}.
Effective masses and effective bag constants for quark matter
have been extensively discussed, e.g.,
within the Nambu--Jona-Lasinio model \cite{Buballa457PLB261}
and within a quasiparticle model \cite{Schertler1997NPA616}.
In principle, not only masses will change in the medium but also the
coupling constant will run in the medium \cite{Fraga2004hepph0412298}.
The question now is how to treat the thermodynamic formulas which
do not violate the fundamental principles of thermodynamics,
when introducing density and/or temperature dependent masses.
In fact, a lot of problems have been caused in this field.

There exist in literature several kinds of thermodynamic treatments.
The first one uses all the thermodynamic formulas exactly
the same as in the constant-mass case
\cite{Chakrabarty1989PLB229,Cleymans1994ZPC62,Letessier1994PLB323}.
The second treatment adds a new term, originated from density
dependence of quark masses, to both pressure and energy
\cite{Benvenuto1989PRD51}. These two treatments were later
proved to be inconsistent with the necessary thermodynamic
requirement: their free energy minimum do not correspond to zero
pressure \cite{Peng2000PRC62}.
The third treatment adds the term from the density dependence of
quark masses to the pressure but not to the energy, and the inconsistency
disappear \cite{Peng2000PRC62}.
Another treatment is the addition of a new term to the thermodynamic
potential. This has been done when masses depend on either
temperature \cite{Gorenstein1995PRD52}
or chemical potential/density
\cite{Schertler1997NPA616,Wangp2000PRC62}.
However, if particle masses depend on both chemical/density
and temperature, this way meets difficulties. We will discuss it
 further in the next section.

With the third thermodynamic treatment  
and the cubic root scaling \cite{Peng2000PRC61},
Zheng {\sl et al}.\ have studied the viscosity of SQM and calculated
the damping time scale due to the coupling of the viscosity
and $r$ mode \cite{Zhengxp2004PRC70}. This model has also been
applied to investigating the quark-diquark equation of state
and compact star structure \cite{Lugonese2003IJMPD12}.

Another important progress has been made recently by
Zhang {\sl et al.}\
\cite{Zhangy2002PRC65,Zhangy2001EPL56,Zhangy2003PRC67,Zhangy2003MPLA18}.
They extended the quark mass density dependent model to finite temperature
to let the model be able to describe phase transition.
In this case, the quark masses depend on both density and temperature
and the permanent confinement is removed.
They use the third thermodynamic treatment mentioned above \cite{Peng2000PRC62}
and parametrize the common interacting part of quark masses
as \cite{Zhangy2002PRC65}
$ 
m_{\mathrm{I}}=B_0(3n_{\mathrm{b}})^{-1}
[1-\left(T/T_c\right)^2]
$ 
with $T_{\mathrm{c}}$ being the critical temperature.
Later, they found this parametrization causes an unreasonable result:
the radius of strangelets decreases with increasing temperature
\cite{Zhangy2001EPL56}.
So they added a new linear term, and the parametrization became
\cite{Zhangy2001EPL56,Zhangy2003PRC67}
\begin{equation} \label{Sumass2}
m_{\mathrm{I}}=\frac{B_0}{3n_{\mathrm{B}}}
\left[1-a\frac{T}{T_c}+b\left(\frac{T}{T_c}\right)^2\right].
\end{equation}

This extension of the model has soon been applied to the study of
strangelets \cite{Zhangy2001EPL56,Zhangy2003PRC67,Zhangy2003MPLA18},
dibaryons \cite{Zhangy2004JPG30},
and proto strange stars \cite{Gupta2004IJMPD12}, etc..

The purpose of the present paper is two-folded.
First, we would like to point out that the thermodynamic derivation
in Ref.\ \cite{Peng2000PRC62} is mainly concentrated on density
dependent masses.
When masses are also temperature dependent, there are special issues
to be considered. We will prove that the temperature dependence of
quark masses causes another term which contributes to the entropy
and energy.
With the new term, originated from the temperature dependence of masses,
added to the entropy,  the quark mass scaling
Eq.\ (\ref{Sumass2}) has a serious problem, i.e.\
$\lim_{T\rightarrow 0}\partial m_{\mathrm{I}}/\partial T\neq 0$
which leads to a non-zero entropy at zero temperature, violating
the third law of thermodynamics. Therefore, our second purpose is to
derive or suggest a new quark mass scaling, and then study the
properties of bulk SQM and strangelets. An interesting new observation
is that low mass strangelets near $\beta$ equilibrium are multiquark
states with an anti-strange quark, such as
the pentaquark $(u^2d^2\bar{s})$ for baryon number 1
and octaquark $(u^4d^3\bar{s})$ for dibaryon etc.

The paper is organized as follows.
In Sec.\ \ref{thermo}, we derive, in detail,
the thermodynamic formulas at finite temperature with density and/or
temperature dependent masses, and show why, in the present version
of the quark mass density and temperature dependent model,
one should require
\begin{equation}
\lim_{T\rightarrow 0}\frac{\partial m_{\mathrm{q}}}{\partial T}=0.
\end{equation}
In the subsequent Sec.\ \ref{qmass}, a new quark mass scaling at
finite temperature is derived or suggested based on chiral and
string model arguments, and accordingly in Sec.\ \ref{bSQM}
and Sec.\ \ref{slets}, respectively, the thermodynamic properties
of bulk SQM and strangelets at both zero and finite temperature
are calculated with the new thermodynamic formulas and the new
quark mass scaling. Finally, a summary is given in Sec.\ \ref{sum}.

\section{Thermodynamics with density and temperature dependent masses}
\label{thermo}

Quasi particle models have been explored in great detail over the
past 10-15 years \cite{Peshier2002PRD66}. It is well understood
how to construct thermodynamically consistent models when
the masses depend on the chemical potential and temperature.
In the present model, however, there are three differences.
First, the particle masses depend on density and temperature,
not directly on chemical potential and temperature. Secondly,
the density and temperature dependence is independently
determined from other arguments. And thirdly, the finite
size effects have to be included. Therefore,
we derive, in the following, the thermodynamic formulas
suitable for the present case.

Suppose the thermodynamic potential is known as
a function of the temperature $T$, volume $V$, chemical potentials
$\{\mu_i\}$, and particle masses $\{m_i\}$, i.e.,
\begin{equation}  \label{Ommod}
\bar{\Omega}=\bar{\Omega}(T,\{\mu_i\},\{m_i\},V).
\end{equation}

  We here explicitly write out the arguments to
make the meaning of partial derivatives clear in the following.
  If the masses $m_i$ are constant, other thermodynamic quantities
can be obtained from normal formulas available in textbooks.
Here the quark masses are density and/or temperature dependent, i.e.,
\begin{equation}
m_i=m_i(n_{\mathrm{b}},T),
\end{equation}
where the baryon number density $n_{\mathrm{b}}$ is connected to
the particle numbers $\{\bar{N}_i\}$ and volume $V$ by
\begin{equation}
n_{\mathrm{b}} = \sum_i n_i/3
\ \ \mbox{with} \ \
 n_i\equiv \sum_i\bar{N}_i/V.
\end{equation}

To study this question, we start from the fundamental derivation equality
of thermodynamics, i.e.,
\begin{equation} \label{basE}
d\bar{E}=Td\bar{S}-PdV+\sum_i\mu_id\bar{N}_i.
\end{equation}
This is nothing but the combination of the first and second laws of
thermodynamics. It means that the energy $\bar{E}$ is the characteristic
function, i.e., all other quantities can be obtained from it,
if one takes the entropy $\bar{S}$, the volume $V$, and the particle
numbers $\{\bar{N}_i\}$ as the full independent state variables.
But it is sometimes convenient to
take $(T,V,\{\mu_i\})$ as the full independent state variables.
In this case, the characteristic function is the thermodynamic potential
$\bar{\Omega}$ which is defined to be
\begin{equation} \label{Omdef}
\bar{\Omega}\equiv\bar{E}-T\bar{S}-\sum_i\mu_i\bar{N}_i
\end{equation}
because adding $-d(T\bar{S}+\sum_i\mu_iN_i)$ to both side of
Eq.\ (\ref{basE}) will give
\begin{equation}
d\bar{\Omega}=-\bar{S}dT-PdV-\sum_i\bar{N}_id\mu_i.
\end{equation}

Another important characteristic function is
the free energy $\bar{F}$. It is defined to be
\begin{equation}  \label{Fdef}
\bar{F}\equiv \bar{E}-T\bar{S}.
\end{equation} \label{Omegadef}
Then the corresponding basic derivation equation is
\begin{equation} \label{basF}
d\bar{F}=-\bar{S}dT-PdV+\sum_i\mu_id\bar{N}_i
\end{equation}
which can be obtained by adding $-d(T\bar{S})$ to Eq.\ (\ref{basE}).
Therefore, the independent state variables are $(T,V,\{\bar{N}\})$
in this case, i.e.,
\begin{equation}
\bar{F}=\bar{F}(T,V,\{N_i\}).
\end{equation}
According to the second term on the right hand side of Eq.\ (\ref{basF}), one has
a general expression for the pressure
\begin{equation} \label{Ptmp1}
P=-\left.\frac{dF}{dV}\right|_{T,\{N_i\}}.
\end{equation}
Here $\bar{F}$ should be expressed as a function of $(T,V,\{\bar{N}_i\})$,
and the derivative is taken with respect to the volume
at fixed $T$ and $\{\bar{N}_i\}$. Comparing Eqs.\ (\ref{Omdef}) and (\ref{Fdef})
leads to the basic relation between thermodynamics and statistics, i.e.,
\begin{equation} \label{Fbargen}
\bar{F}=\bar{\Omega}+\sum_i\mu_i\bar{N}_i
\end{equation}
Substituting this into Eq.\ (\ref{Ptmp1}) gives
\begin{equation} \label{Ptmp2}
P =-\left.\frac{d\bar{\Omega}}{dV}\right|_{T,\{N_i\}}
    -\sum_i \bar{N}_i\frac{\partial\mu_i}{\partial V}.
\end{equation}
Because the independent state variables here are $(T,V,\{N_i\})$,
the chemicals $\{\mu_i\}$ in $\bar{\Omega}$ [see Eq.\ (\ref{Ommod})]
should be expressed as a function of $(T,V,\{N_i\})$, i.e.,
\begin{equation} \label{mufunc}
\mu_i=\mu_i(T,V,\{N_k\}).
\end{equation}
So the total derivative of $\bar{\Omega}$ with respect to $V$
at fixed $T$ and $\{N_i\}$ is
\begin{equation} \label{totdOdV}
\left.\frac{d\bar{\Omega}}{dV}\right|_{T,\{N_k\}}
=\frac{\partial\bar{\Omega}}{\partial V}
 +\sum_i\frac{\partial\bar{\Omega}}{\partial\mu_i}
         \frac{\partial\mu_i}{\partial V}
 +\sum_i\frac{\partial\bar{\Omega}}{\partial m_i}
         \frac{\partial m_i}{\partial V},
\end{equation}
where
\begin{equation} \label{dmdVexpl}
\frac{\partial m_i}{\partial V}
\equiv
\left.\frac{\partial m_i}{\partial V}\right|_{T,\{N_k\}}
=\frac{\partial m_i}{\partial n_{\mathrm{b}}}
 \frac{\partial n_{\mathrm{b}}}{\partial V}
=-\frac{n_{\mathrm{b}}}{V}\frac{\partial m_i}{\partial n_{\mathrm{b}}}.
\end{equation}
Consequently, substitution of Eq.\ (\ref{totdOdV})
into Eq.\ (\ref{Ptmp1}) gives
\begin{equation}  \label{Pbargen}
P
=-\frac{\partial\bar{\Omega}}{\partial V}
   +\frac{n_{\mathrm{b}}}{V}
    \sum_i\frac{\partial\bar{\Omega}}{\partial m_i}
    \frac{\partial m_i}{\partial n_{\mathrm{b}}}
 -\sum_i\left(
          \bar{N}_i+\frac{\partial\bar{\Omega}}{\partial\mu_i}
        \right)
        \frac{\partial\mu_i}{\partial V}.
\end{equation}

Similarly, for the entropy, we have
\begin{equation}
\bar{S}
=-\left.\frac{d\bar{F}}{dT}\right|_{V,\{N_k\}}
=-\left.\frac{d\bar{\Omega}}{dT}\right|_{V,\{N_k\}}
 -\sum_i\bar{N}_i\frac{\partial\mu_i}{\partial T}.
\end{equation}
Substitution of
\begin{equation}
\left.\frac{d\bar{\Omega}}{dT}\right|_{V,\{N_k\}}
= \frac{\partial\bar{\Omega}}{\partial T}
 +\sum_i\frac{\partial\bar{\Omega}}{\partial\mu_i}
        \frac{\partial\mu_i}{\partial T}
 +\sum_i\frac{\partial\bar{\Omega}}{\partial m_i}
        \frac{\partial m_i}{\partial T}
\end{equation}
leads to
\begin{equation}  \label{Sbargen}
\bar{S}
 =-\frac{\partial\bar{\Omega}}{\partial T}
   -\sum_i\frac{\partial\bar{\Omega}}{\partial m_i}
    \frac{\partial m_i}{\partial T}
   -\sum_i\left(
            \bar{N}_i+\frac{\partial\bar{\Omega}}{\partial\mu_i}
           \right)
           \frac{\partial\mu_i}{\partial T}.
\end{equation}

The energy can be obtained from Eq.\ (\ref{Omdef})
\begin{equation}
\bar{E}=\bar{\Omega}+\sum_i\mu_i\bar{N}_i+T\bar{S}.
\end{equation}
Replacing $\bar{S}$ here with the expression in Eq.\ (\ref{Sbargen}),
one has
\begin{eqnarray} \label{Ebexpgen}
\bar{E}
&=&
    \bar{\Omega}
   +\sum_i\mu_i \bar{N}_i
   -T\frac{\partial\bar{\Omega}}{\partial T}
   -T\sum_i\frac{\partial\bar{\Omega}}{\partial m_i}
         \frac{\partial m_i}{\partial T}
                  \nonumber\\
 & &
      \label{Ebargen}
   -T\sum_i\left(
             \bar{N}_i+\frac{\partial\bar{\Omega}}{\partial\mu_i}
           \right)
           \frac{\partial \mu_i}{\partial T}.
 \end{eqnarray}

Now we need $N_{\mathrm{f}}$ (number of flavors) equations to connect
$T$, $V$, $\{\mu_i\}$, and $\bar{N}_i$, so that the functions
$\mu_i(T,V,\{N_k\})$ in Eq.\ (\ref{mufunc}) can be obtained.
Presently, nearly all relevant models adopt
\cite{Chakrabarty1989PLB229,Benvenuto1989PRD51,Peng2000PRC62,%
Gorenstein1995PRD52,Schertler1997NPA616,Wangp2000PRC62,%
Zhangy2002PRC65,Zhangy2003MPLA18,Zhangy2003PRC67,Zhangy2004JPG30,%
Gupta2004IJMPD12}
\begin{equation} \label{NiMDTD}
\bar{N}_i
=-\left.
 \frac{\partial\bar{\Omega}}{\partial\mu_i}
  \right|_{T,V,\{m_k\}}.
\end{equation}
Please note, $\bar{N}_i$ also appears on the right hand side of this equation
through $m_k=m_k(\sum_i\bar{N}_i/[3V],T)$. Then $\mu_i$\ can be
solved as a function of $T$, $V$, and $\{N_k\}$ from these equations.

If Eq.\ (\ref{NiMDTD}) applies, the last term in
Eqs.\ (\ref{Pbargen}), (\ref{Sbargen}), and (\ref{Ebargen}) vanishes.
So all the formulas will take the simplest form.
If define $\Omega\equiv\bar{\Omega}/V$, $E\equiv\bar{E}/V$,
$S\equiv\bar{S}/V$, $n_i\equiv \bar{N}_i/V$, and use $V=(4/3)\pi R^3$,
then Eqs.\ (\ref{Pbargen}), (\ref{Ebargen}), (\ref{Sbargen}), (\ref{Fbargen}),
and Eq.\ (\ref{NiMDTD}) become, respectively,
\begin{eqnarray}
P&=& \label{PMDTD}
  -\Omega
  -V\frac{\partial\Omega}{\partial V}
  +n_{\mathrm{b}}
     \sum_i\frac{\partial\Omega}{\partial m_i}
           \frac{\partial m_i}{\partial n_{\mathrm{b}}}, \\
E&=&   \label{EMDTD}
  \Omega-\sum_i\mu_i\frac{\partial\Omega}{\partial\mu_i}
               -T\frac{\partial\Omega}{\partial T}
  -T\sum_i\frac{\partial\Omega}{\partial m_i}
         \frac{\partial m_i}{\partial T},  \\
S&=&  \label{SMDTD}
   -\frac{\partial\Omega}{\partial T}
   -\sum_i\frac{\partial\Omega}{\partial m_i}
    \frac{\partial m_i}{\partial T},  \label{Sfinexp} \\
F&=&   \label{Fgenshort}
   \Omega-\sum_i\mu_i\frac{\partial\Omega}{\partial\mu_i},\\
n_i&=& -\frac{\partial\Omega}{\partial\mu_i}. \label{nMDTD}
\end{eqnarray}

Compared with thermodynamic formulas in the normal case,
both the entropy and energy have a new term from the
temperature dependence of the masses, while the pressure
has a new term due to the density dependence of the masses.
The second term in Eq.\ (\ref{PMDTD}) exists
when the finite size effect can not be ignored, no matter
the masses are constant or not.
These new terms are understandable with a view to the equalities
\begin{equation}
S=-\left.\frac{d\bar{\Omega}}{dT}\right|_{V,\mu},
\ \ \ \
P= -\left.\frac{d\bar{\Omega}}{dV}\right|_{T,\mu},
\end{equation}
and the total derivative rules in mathematics.
Here one should pay special attention to the difference between
the total derivatives and the partial derivatives
in Eq.\ (\ref{Pbargen}) and Eq.\ (\ref{Sbargen}).

>From the viewpoint of quasiparticle models,
the first and second terms in the pressure Eq.\ (\ref{PMDTD})
are the normal quasiparticle contributions, while the last
extra term is the contribution of mean-field interactions.

Normally, the first term on the right hand side of Eq.\ (\ref{Sfinexp})
goes to zero when the temperature approaches to zero, i.e.,
$\lim_{T\rightarrow 0}\partial\Omega/\partial T=0$.
However, the first factor of the second term does not, i.e.,
$\lim_{T\rightarrow 0}\partial\Omega/\partial m_i \neq 0$.
Therefore, one must require
\begin{equation} \label{limmT0}
\lim_{T\rightarrow 0}
\frac{\partial m_i}{\partial T} = 0
\end{equation}
to be consistent with the third law of thermodynamics, i.e.,
$\lim_{T\rightarrow 0}S=0$.
Because Eq.\ (\ref{SMDTD}) depends on the model
assumption Eq.\ (\ref{NiMDTD}), Eq.\ (\ref{limmT0})
is a model dependent requirement.
Consequently, the thermodynamic formulas
Eqs.\ (\ref{PMDTD})-(\ref{nMDTD}) are merely valid for
systems whose interactions meet this requirement.
In the subsequent section, we will see that Eq.\ (\ref{limmT0}) is
indeed fulfilled by a new quark mass scaling
for the strong interactions of quarks.

In Ref.\ \cite{Peng2000PRC62}, we mentioned a necessary condition
any consistent thermodynamic treatment must satisfy.
At finite temperature, we also have a similar criterion:
\begin{equation} \label{pnb}
P= -\left.\frac{d\bar{F}}{dn_{\mathrm{b}}}\right|_{T,\{\bar{N}_k\}}
   \frac{\partial n_{\mathrm{b}}}{\partial V}
  = n_{\mathrm{b}}^2 \frac{d}{dn_{\mathrm{b}}}
    \left(
     \frac{F}{n_{\mathrm{b}}}
    \right)_{T,\{\bar{N}_k\}}.
\end{equation}
In obtaining the second equality of Eq.\ (\ref{pnb}), the equalities
$V=\sum_i\bar{N}_i/(3n_{\mathrm{b}})$,
$\partial n_{\mathrm{b}}/\partial V=-n_{\mathrm{b}}/V$,
and $\bar{F}=VF$ have been used.
Because $F/n_{\mathrm{b}}=\bar{F}/(\sum_i\bar{N}_i/3)$
is the free energy per baryon, Eq.\ (\ref{pnb}) shows explicitly that
the free energy extreme occurs exactly at zero pressure.
This is why people are interested in the free energy minimum,
rather than the energy minimum, at finite temperature,
to look for mechanically stable states.

The extra term or the last term in Eq.\ (\ref{PMDTD})
is important even in the MIT bag model when one introduces
a density dependent bag constant $B(n_{\mathrm{b}})$.
In this context, the extra term is $n_{\mathrm{b}}dB/dn_{\mathrm{b}}$.
If it is not included, the zero pressure will not be located
at the free energy minimum. In Ref.\ \cite{Burgio2002PRC66}, this term
has been considered in the calculation of hadron-quark phase transition
in dense matter and neutron stars within the bag model.

The derivation process of the pressure is also instructive
to the case when one wants to include the Coulomb contribution.
In this case, the energy density, accordingly the thermodynamic potential
density, gets a term $E_{\mathrm{Coul}}(\{n_k\},R)$.
Correspondingly the pressure gets
\begin{equation}
P_{\mathrm{Coul}}
=-E_{\mathrm{Coul}}
 +\sum_in_i\frac{\partial E_{\mathrm{Coul}}}{\partial n_i}
 -\frac{R}{3}\frac{\partial E_{\mathrm{Coul}}}{\partial R}.
\end{equation}

In literature, there is another approach to have thermodynamic
consistency by adding a term $B^*$ to the original
thermodynamic potential density. This makes the total thermodynamic potential
density becomes $\Omega+B^*$
\cite{Gorenstein1995PRD52,Schertler1997NPA616,Wangp2000PRC62}.
We comment that this can be unconditionally done merely in two
special cases, i.e., at finite temperature with zero chemical
potential and at finite chemical potential with zero temperature.
The expression of the added term for the former case is
 \cite{Gorenstein1995PRD52}:
\begin{equation} \label{Bstar1}
B^*(T)
=-\int_{T_0}^T\left.
   \frac{\partial\Omega}{\partial m}
      \right|_{T,\mu=0}
 \frac{dm}{dT} dT,
\end{equation}
while that for the later case is
\cite{Schertler1997NPA616}:
\begin{equation} \label{Bstar2}
B^*(\mu)
=-\int_{\mu_0}^{\mu}\left.
 \frac{\partial\Omega}{\partial m}
                    \right|_{T=0,\mu}
 \frac{dm}{d\mu} d\mu.
\end{equation}

However, if one wants to extend this to both chemical potential/density
and temperature dependent masses, difficulties arise.
To perform the integration in Eqs.\ (\ref{Bstar1}) and (\ref{Bstar2}),
one should use $m=m(T)$ to replace the $m$ on the right hand
side of Eq.\ (\ref{Bstar1}) or apply $m=m(\mu)$ to
the right hand side of Eq.\ (\ref{Bstar2}).
If $m_i=m_i(T,\{\mu_k\})$,
the directly extended integration is
\begin{equation} \label{Bstar3}
B^*(T,\mu)
=-\int 
  \left[
  \sum_i\frac{\partial\Omega}{\partial m_i}
        \frac{\partial m_i}{\partial T} dT
 +\sum_{i,j}\frac{\partial\Omega}{\partial m_i}
        \frac{\partial m_i}{\partial \mu_j} d\mu_j
      \right].
\end{equation}
In the above, ($T_0, \mu_0$) is some reference point.
Eq.\ (\ref{Bstar3}) is a multi-dimensional integration.
To let it be path-independent, one must  mathematically
require Cauchy conditions. If all chemical potentials are
equal, for example, the Cauchy condition is
\begin{equation} \label{Causheq}
\sum_i\left[
\frac{\partial^2\Omega}{\partial m_i\partial\mu}
\frac{\partial m_i}{\partial T}
-\frac{\partial^2\Omega}{\partial m_i\partial T}
\frac{\partial m_i}{\partial\mu}
       \right]=0.
\end{equation}
Such an example has recently been given in
Ref.\ \cite{Peshier2000PRC61}, where the
Cauchy condition Eq.\ (\ref{Causheq}), or the equivalent
Maxwell relation Eq.\ (7) in Ref.\ \cite{Peshier2000PRC61},
is fulfilled by solving the equation for the coupling in the masses,
As the Eq.\ (8) of Ref.\ \cite{Peshier2000PRC61} indicated.
However, if the masses are completely determined from other
arguments, and so there are no adjustable parameters, this
thermodynamic treatment may fail.
When masses depend on density and temperature, rather than
directly on chemical potential and temperature, the case
becomes much more involved.
Also, if finite size effects can not be ignored, or in other words,
$\Omega$ depends explicitly on the volume or radius,
not merely the integration in Eq.\ (\ref{Bstar3}) becomes
much more difficult or impossible, the integration
in Eq.\ (\ref{Bstar1}) or in Eq.\ (\ref{Bstar2})
has an unknown function of the volume or radius as well.

These difficulties are not surprising.
In fact, we have proved, from Eq.\ (\ref{basE}) to Eq.\ (\ref{Ebexpgen}),
that Eqs.\ (\ref{PMDTD})-(\ref{Fgenshort}) are inevitable consequences
of Eq.\ (\ref{Ommod}) with Eq.\ (\ref{NiMDTD}).
In this paper, we will apply these formulas to the calculation of
properties of both bulk SQM and strangelets.

\section{Derivation of quark mass scaling}
\label{qmass}

In the preceding section, we have derived the thermodynamic formulas
suitable for systems with density and/or temperature dependent masses.
In this section, we derive quark mass scaling by a similar method
as in Ref.\ \cite{Peng2000PRC61}.

Let's schematically write the QCD hamiltonian density
for the three flavor case as
\begin{equation}
H_{\mathrm{QCD}}=H_{\mathrm{k}}+\sum_{q=u,d,s}m_{q0}\bar{q}q
               +H_{\mathrm{I}},
\end{equation}
where $m_{q0}$ (q=u,d,s) are the quark's current mass,
$H_{\mathrm{k}}$ is the kinetic term, $H_{\mathrm{I}}$ is the
interacting part.

Now we want to include interaction effects within an
equivalent mass $m_{\mathrm{q}}$. For this purpose we
define an hamiltonian density of the form
\begin{equation}
H_{\mathrm{eqv}}=H_{\mathrm{k}}+\sum\limits_{q=u,d,s}m_q\bar{q}q,
\end{equation}
where $m_q$ is our equivalent mass to be determined.
We firstly divide it into two parts, i.e.\
\begin{equation} \label{mqm0mI}
m_q=m_{q0}+m_{\mathrm{I}}.
\end{equation}
The first part $m_{q0}$ $(q=u,d,s)$ are the quark current masses
while $m_{\mathrm{I}}$ is a common part for all the three flavors
to mimic the strong interaction.
Obviously we must require that the two hamiltonian densities
$H_{\mathrm{eqv}}$ and $H_{\mathrm{QCD}}$
have the same eigenenergy for any eigenstate $|\Psi\rangle$,
i.e.\
\begin{equation}
\langle\Psi|H_{\mathrm{eqv}}|\Psi\rangle
= \langle\Psi|H_{\mathrm{QCD}}|\Psi\rangle.
\end{equation}

Applying this equality, respectively, to the state
 $|n_{\mathrm{b}},T\rangle$\ and the vacuum $|0\rangle$,
and then taking the difference, we have
\begin{equation}
\langle H_{\mathrm{eqv}}\rangle_{n_{\mathrm{b}},T}
  -\langle H_{\mathrm{eqv}}\rangle_{0},
=\langle H_{\mathrm{QCD}}\rangle_{n_{\mathrm{b}},T}
  -\langle H_{\mathrm{QCD}}\rangle_{0},
\end{equation}
where the symbol definitions
$\langle O\rangle_{n_{\mathrm{b}},T}
\equiv \langle n_{\mathrm{b}},T|O|n_{\mathrm{b}},T\rangle$
and $\langle O\rangle_0 \equiv\langle 0|O|0\rangle$
have been used for an arbitrary  operator $O$.
Then solving for $m_{\mathrm{I}}$ from this equation gives
\begin{equation} \label{mI}
m_{\mathrm{I}}
=\frac{ \langle H_{\mathrm{I}}\rangle }
{\sum\limits_{q=u,d,s}\left[\langle\bar{q}q\rangle_{n_{\mathrm{b}},T}
            -\langle\bar{q}q\rangle_0\right]},
\end{equation}
where
$
\langle H_{\mathrm{I}}\rangle
\equiv
 \langle H_{\mathrm{I}}\rangle_{n_{\mathrm{b}},T}
-\langle H_{\mathrm{I}}\rangle_0
$
is the interacting part of the energy density from strong
interactions between quarks. It can be linked to density $n_{\mathrm{b}}$
and temperature $T$ by
\begin{equation} \label{HirT}
\langle H_{\mathrm{I}}\rangle=3n_{\mathrm{b}} \mathrm{v}(\bar{r},T).
\end{equation}
Here
\begin{equation}
 \bar{r}=\left(\frac{2}{\pi n_{\mathrm{b}}}\right)^{1/3}
\end{equation}
is the average distance of quarks at the density $n_{\mathrm{b}}$,
$\mbox{v}(\bar{r},T)$ is the interaction between quarks at density
$n_{\mathrm{b}}$ and temperature $T$. Because we are interested in
the confinement effect,
while the lattice simulation \cite{Belyaev1984PLB136}
and string model investigation \cite{Isgur1983PLB124}
show that the confinement is linear, we write
\begin{equation} \label{sigr}
\mbox{v}(n_{\mathrm{b}},T)=\sigma(T)\bar{r}.
\end{equation}

  The temperature dependence of the string tension $\sigma(T)$
can be obtained by combining the Eqs.\ (94) and (91) in Ref.\
\cite{Ukawa}:
\begin{equation} \label{sig1}
\sigma(T)=\sigma_0-\frac{4T}{a}\exp\left(-\frac{2\sigma_0a}{T}\right),
\end{equation}
where $a$ is the lattice spacing while $\sigma_0$ is the
string tension at zero temperature. The value of $\sigma_0$
from potential models varies in the range of
(0.18, 0.22) GeV$^2$ \cite{Veseli}.

For convenience, let's define a dimensionless constant
 $\lambda\equiv 2\sigma_0a/T_{\mathrm{c}}$ with $T_c$ being the
critical temperature.  Then substituting
 $a=\lambda T_{\mathrm{c}}/(2\sigma_0)$ into Eq.\ (\ref{sig1}) gives
\begin{equation}
\sigma(T)=\sigma_0 \left[1-\frac{8T}{\lambda T_{\mathrm{c}}}
                \exp\left(-\lambda\frac{T_{\mathrm{c}}}{T}\right)
                \right].
\end{equation}

  Because the string tension should become zero at the
deconfinement temperature, the value of $\lambda$ is determined
by the equation $\sigma(T_{\mathrm{c}})=0$ whose solution is
\begin{equation}
\lambda=\mbox{LambertW}(8)\approx 1.60581199632.
\end{equation}

Accordingly, Eq.\ (\ref{sigr}) becomes
\begin{equation} \label{vnbT}
\mbox{v}(n_{\mathrm{b}},T)
  =\frac{(2/\pi)^{1/3}\sigma_0}{n_{\mathrm{b}}^{1/3}}
 \left(1-\frac{8T}{\lambda T_{\mathrm{c}}}
                e^{-\lambda T_{\mathrm{c}}/T}
                \right).
\end{equation}

\begin{figure}
\epsfxsize=8.2cm
\epsfbox{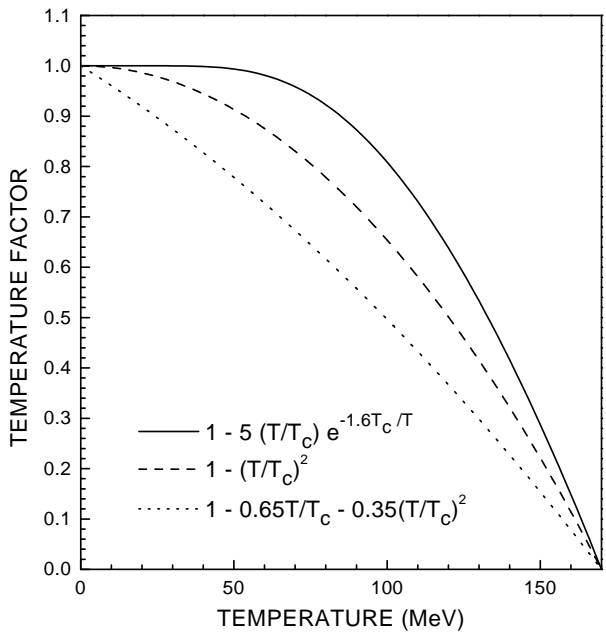}
\caption{
  Comparison of different temperature factors.
  The full line, dashed line, and dotted line are, respectively, for
  $1-\frac{8T}{\lambda T_\mathrm{c}} e^{-\lambda T_{\mathrm{c}}/T}$,
  $1-\left(\frac{T}{T_{\mathrm{c}}}\right)^2$,
  and $1-0.65\frac{T}{T_{\mathrm{c}}}
        -0.35\left(\frac{T}{T_{\mathrm{c}}}\right)^2$, where
        $T_{\mathrm{c}}$ is the critical
        temperature and $\lambda=\mathrm{LambertW}(8)\approx 1.6$.
         }
\label{pot}
\end{figure}

The inter-quark potential has been also studied by
comparison to lattice data in Ref.\ \cite{Wong65PRC034902}.
Replacing the factor $\exp(-\mu r)$ in the Eq.\ (2.8) there
with $1-\mu r$, one will find the linear confining part.
The temperature factor used there is $1-(T/T_{\mathrm{c}})^2$,
and this temperature factor has also been used in Ref.\
\cite{Zhangy2002PRC65}. However, because of a problem with the
radius-temperature relation mentioned in the introduction part,
the factor has been modified to
$1-0.65T/T_{\mathrm{c}}-0.35(T/T_{\mathrm{c}})^2$
\cite{Zhangy2001EPL56,Zhangy2003PRC67,Zhangy2003MPLA18,Zhangy2004JPG30}.
The temperature factor derived in the present paper is
$
1-\frac{8T}{\lambda T_{\mathrm{c}}}
         e^{-\lambda T_{\mathrm{c}}/T}.
$
These temperature factors are compared in Fig.\ \ref{pot}.

Chiral condensates have been extensively studied in literature
\cite{qcrev,Peng747NPA75}.
In principle, they depend on both density and temperature.
Presently for simplicity, we only consider their density dependence
and use the model-independent result \cite{Drukarev353ZPA455,Cohen1991PRL67}
\begin{equation}  \label{qclin}
\frac{\langle\bar{q}q\rangle_{n_{\mathrm{b}}}}{\langle\bar{q}q\rangle_0}
=1-\frac{n_{\mathrm{b}}}{\rho^*},
\end{equation}
with
\begin{equation}
\rho^*=\frac{m_{\pi}^2f_{\pi}^2}{\sigma_{\mathrm{N}}}.
\end{equation}
When taking 140 MeV for the pion mass $m_{\pi}$, 93.3 MeV
for the pion decay constant $f_{\pi}$,
and 45 MeV for pion-nucleon sigma term $\sigma_{\mathrm{N}}$,
one has $\rho^*\approx$\ 0.49 fm$^{-3}$.

  Substituting Eqs.\ (\ref{qclin}), (\ref{vnbT}), and (\ref{HirT})
to Eq.\ (\ref{mI}), we obtain
\begin{equation} \label{mIfin}
m_{\mathrm{I}}(n_{\mathrm{b}},T)=\frac{D}{n_{\mathrm{b}}^z}
  \left[
1-\frac{8T}{\lambda T_c}
 \exp\left(-\lambda\frac{T_{\mathrm{c}}}{T}\right)
  \right].
\end{equation}
Here $z=1/3$.  
Please note, many density and temperature independent
constants such as the vacuum condensates, the string tension
at zero temperature, and $\rho^*$ et al.\ are grouped into
a constant $D$, i.e.,
\begin{equation} \label{Dref}
D=\frac{3(2/\pi)^{1/3}\sigma_0\rho^*}
       {-\sum_q\langle\bar{q}q\rangle_0}.
\end{equation}
Taking $\sigma_0=0.18$ GeV$^2$ and $\rho^*=0.49$ fm$^{-3}$,
$\sqrt{D}$ value is in the range of $(147, 270)$ MeV when
$-\sum_q\langle\bar{q}q\rangle_0$
varies from 3$\times$(300 MeV)$^3$ to 3$\times$(200 MeV)$^3$.
Becasue of many uncertainties within the
relevant quantities and the fact that
Eq.\ (\ref{qclin}) is not exactly valid, at least it
ignores temperature dependence and higher orders in density,
we do not try to use the relevant quantities to
calculate the value for $D$ from Eq.\ (\ref{Dref}).
Instead, we treat $D$
as a free parameter to be determined by stability arguments,
i.e., it makes the energy per baryon $E/n_{\mathrm{b}}$
at zero temperature is greater than 930 MeV for two flavor
quark matter in order not to contradict standard nuclear physics,
but less than 930 MeV for three flavor quark matter so that
SQM can have a chance to be absolutely stable.
Obviously, the range determined by this method depends on
the thermodynamic formulas and the values of quark current masses.
In the present calculation, we use
 $m_{u0}=5$ MeV,  $m_{d0}=10$ MeV, and $m_{s0}=120$ MeV
for the current masses involved. These conditions constrain $\sqrt{D}$
to a very narrow range of (154.8278, 156.1655) MeV,
and we take $D=(156\ \mbox{MeV})^2$ in this paper.

The model described in the above is a combination of a
chiral model and a string model. There should be nothing
really wrong with it. In fact, its zero-temperature form
has been successfully applied to studying the properties of
SQM \cite{Peng2000PRC62,Peng2000PRC61},
calculating the damping time scale of strange stars due to
the coupling of the viscosity and r mode \cite{Zhengxp2004PRC70},
and investigating the quark di-quark equation of state
and compact star structure \cite{Lugonese2003IJMPD12}.
Furthermore, various applications of the conventional
quasi particle models with chemical and temperature
dependent masses have turned out to be very successful
 \cite{Peshier2002PRD66,Peshier2000PRC61}.
Therefore, we will apply the specific model presented
here to the investigation of SQM in bulk and strangelets
in the following.

\section{Properties of bulk strange quark matter
         at finite temperature}
\label{bSQM}

As usually done in this model, the quasiparticle contribution to
the total thermodynamic potential density of SQM is written as
\begin{equation} \label{Otmum}
\Omega=\sum\limits_i\Omega_i(T,\mu_i,m_i).
\end{equation}
where the summation index $i$ goes over
$u$, $d$, $s$ quarks and electrons.
Anti-particles are treated as a whole with particles, i.e.,
the contribution of the particle type $i$ to the thermodynamic
potential density is
\begin{eqnarray} \label{Omegaint}
\Omega_i
&=&
   -\frac{g_iT}{2\pi^2}
   \int_0^{\infty}
    \left\{
 \ln\left[1+e^{-(\epsilon_{i,p}-\mu_i)/T}\right]
    \right. \nonumber\\
&&  \left.\phantom{-\frac{g_iT}{2\pi^2}}
+\ln\left[1+e^{-(\epsilon_{i,p}+\mu_i)/T}\right]
    \right\}
  p^2 \mbox{d}p,
\end{eqnarray}
where $m_i$, $\mu_i$, and $T$ are, respectively, the particle masses,
chemical potentials, temperature, and $\epsilon_{i,p}=(p^2+m_i^2)^{1/2}$
is the dispersion relation.
The particle number density corresponding to the particle type $i$
is obtained by $n_i=-\partial\Omega/\partial\mu_i$, giving
\begin{equation} \label{ni56}
n_i
=\frac{g_i}{2\pi^2}
 \int_0^{\infty}
 \left[
   \frac{1}{1+e^{(\epsilon_{i,p}-\mu_i)/T}}
  -\frac{1}{1+e^{(\epsilon_{i,p}+\mu_i)/T}}
 \right]
p^2 \mbox{d}p.
\end{equation}

The energy density is $E=\sum_i E_i(T,\mu_i,m_i)$ with
\begin{eqnarray}  \label{E0def}
E_i
&=&
 \frac{g_i}{2\pi^2}
 \int_0^{\infty}
   \left[
   \frac{\epsilon_{i,p}\ p^2}{1+e^{(\epsilon_{i,p}-\mu_i)/T}}
  +\frac{\epsilon_{i,p}\ p^2}{1+e^{(\epsilon_{i,p}+\mu_i)/T}}
 \right]
   \mbox{d}p
    \nonumber\\
&&
  -T\frac{\partial\Omega_i}{\partial m_i}
    \frac{\partial m_i}{\partial T}.
\end{eqnarray}
The free energy density $F$, the entropy density $S$,
 and the pressure $P$ are, respectively,
\begin{equation}
F=\sum_i F_i
 =\sum_i\left(
\Omega_i+\mu_i n_i
      \right),
\end{equation}
\begin{equation}
S=\sum_i S_i
  =\sum_i\left(
  -\frac{\partial\Omega_i}{\partial T}
  -\frac{\partial\Omega_i}{\partial m_i}
          \frac{\partial m_i}{\partial T}
        \right),
\end{equation}
\begin{equation}
P=\sum_i P_i
 =\sum_i
  \left(
  -\Omega_i
  + n_{\mathrm{b}}\frac{\partial m_i}{\partial n_{\mathrm{b}}}
  \frac{\partial\Omega_i}{\partial m_i}
  \right).
\end{equation}

In the above, the partial derivatives relevant to $\Omega_i$ are
\begin{eqnarray}
\frac{\partial\Omega_i}{\partial m_i}
&=&
   \frac{g_im_i}{2\pi^2}
   \int_0^{\infty}
   \left[
    \frac{1}{1+e^{(\epsilon_{i,p}-\mu_i)/T}}
   \right. \nonumber\\
&& \left. \phantom{\frac{g_im_i}{2\pi^2}\int_0^{\infty}}
  +\frac{1}{1+e^{(\epsilon_{i,p}+\mu_i)/T}}
 \right]
 \frac{p^2 \mbox{d}p}{\epsilon_{i,p}}
\end{eqnarray}
and
\begin{eqnarray}
\frac{\partial\Omega_i}{\partial T}
&=& -\frac{g_i}{2\pi^2}
  \int_0^{\infty}
  \left\{
 \ln[1+e^{-(\epsilon_{i,p}-\mu_i)/T}]
  \right.\nonumber\\
&& \phantom{-\frac{g_i}{2\pi^2}\int_0^{\infty}}
 +\frac{(\epsilon_{i,p}-\mu_i)/T}{1+e^{(\epsilon_{i,p}-\mu_i)/T}}
    \nonumber\\
&&  \phantom{-\frac{g_i}{2\pi^2}\int_0^{\infty}}
+\ln[1+e^{-(\epsilon_{i,p}+\mu_i)/T}]
   \nonumber\\
&& \left. \phantom{-\frac{g_i}{2\pi^2}\int_0^{\infty}}
+\frac{(\epsilon_{i,p}+\mu_i)/T}{1+e^{(\epsilon_{i,p}+\mu_i)/T}}
  \right\}
  p^2\mathrm{d}p.
\end{eqnarray}

The masses of electrons/positrons (0.511 MeV) and
neutrinos/anti-neutrinos (if any) are extremely small.
So they can be treated to be zero.
For massless particles, the relevant integrations in the above
can be carried out to give
\begin{eqnarray}
\Omega_i 
&=& -\frac{g_i}{24}
 \left(
  \frac{\mu_i^4}{\pi^2}
 +2\mu_i^2T^2
 +\frac{7}{15}\pi^2T^4
 \right),    \label{Omegam0}  \\
 n_i 
  &=&       \label{delnim0}
 \frac{g_i}{6}\mu_i
\left(T^2+\frac{\mu_i^2}{\pi^2}\right), \\
S_i 
&=&
\frac{g_i}{6}T
\left(
\mu_i^2+\frac{7}{15}\pi^2T^2
\right), \\
E_i 
&=&\frac{g_i}{8}
   \left( \frac{\mu_i^4}{\pi^2}
         +2\mu_i^2T^2
         +\frac{7}{15}\pi^2T^4
   \right),\\
F_i 
&=&
\frac{g_i}{8}
\left(
\frac{\mu_i^4}{\pi^2}+\frac{2}{3}\mu_i^2T^2-\frac{7}{45}\pi^2T^4
\right).  \label{Fm0}
\end{eqnarray}

At zero temperature, we have the familiar results
\begin{eqnarray}
\Omega_i 
&=&-\frac{g_i}{48\pi^2}
     \left[
|\mu_i|\sqrt{\mu_i^2-m_i^2}\left(2\mu_i^2-5m_i^2\right)
    \right. \nonumber\\
 && \left. \phantom{-\frac{g_i}{48\pi^2}}
  +3m_i^4\ln\frac{|\mu_i|+\sqrt{\mu_i^2-m_i^2}}{m_i}
    \right], \\
n_i 
&=&\frac{g_i\mu_i}{6\pi^2|\mu_i|}
  \left(\mu_i^2-m_i^2\right)^{3/2},    \\
E_i 
 &=& \frac{g_i}{16\pi^2}
     \left[
|\mu_i|\sqrt{\mu_i^2-m_i^2}\left(2\mu_i^2-m_i^2\right)
     \right.  \nonumber\\
 &&  \left. \phantom{\frac{g_i}{16\pi^2}}
     -m_i^4\ln\frac{|\mu_i|+\sqrt{\mu_i^2-m_i^2}}{m_i}
      \right],   \\
\frac{\partial\Omega_i}{\partial m_i} 
&=&
    \frac{g_im_i}{4\pi^2}
      \left[
    |\mu_i|\sqrt{\mu_i^2-m_i^2}
      \right. \nonumber\\
 &&    \left. \phantom{\frac{g_im_i}{4\pi^2}}
    -m_i^2\ln\frac{|\mu_i|+\sqrt{\mu_i^2-m_i^2}}{m_i}
      \right].
\end{eqnarray}

The above formulas show that the number density is an odd
function of the corresponding chemical potential, i.e.,
particle and anti-particle numbers are opposite in sign.
But other quantities, such as the thermodynamic potential,
entropy, energy, and free energy, are all even functions.

\begin{figure}[htb]
\epsfxsize=8.2cm
\epsfbox{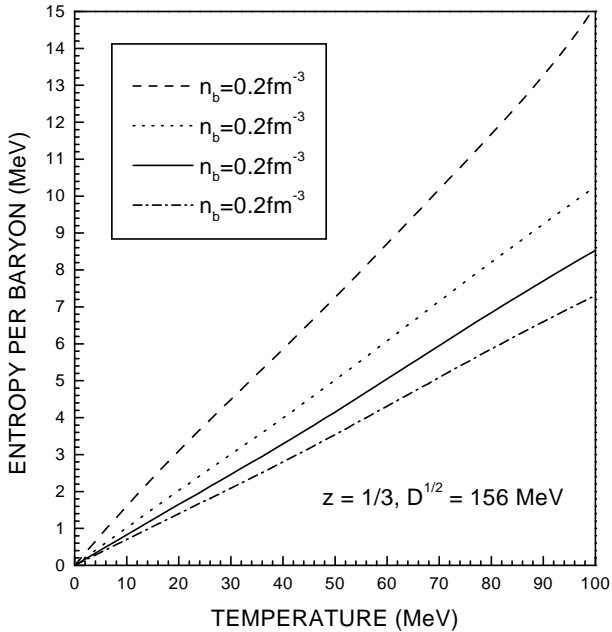}
\caption{
  The entropy per baryon of SQM as an function of temperature.
  It increases with increasing temperature. When the temperature
  approaches to zero, the entropy goes to zero, no matter
  the density is high or low.
         }
\label{S13nb}
\end{figure}

Suppose weak equilibrium is always reached within SQM by
the weak reactions such as
\begin{equation}
d,s\leftrightarrow u+e+\bar{\nu_e}, \ \ s+u \leftrightarrow u+d.
\end{equation}
Correspondingly, relevant chemical potentials satisfy
\begin{eqnarray}
&\mu_d=\mu_s,&    \label{eq1} \\
&\mu_d+\mu_{\nu}=\mu_u+\mu_e.& \label{eq2}
\end{eqnarray}

We also have the baryon number density equality
\begin{equation}
\frac{1}{3}\left(n_u+n_d+n_s\right)=n_{\mathrm{b}}
\end{equation}
and the charge neutrality condition
\begin{equation} \label{chg0}
 \frac{2}{3}n_u
-\frac{1}{3}n_d
-\frac{1}{3}n_s
- n_e=0.
\end{equation}

Neutrinos are assumed to enter and leave the system freely. So
their chemical potential $\mu_{\nu}$\ is zero. From Eqs.\
(\ref{Omegam0})-(\ref{Fm0}), they have no contribution
at zero temperature, but contribute at finite temperature.

For a given $T$ and $n_{\mathrm{b}}$,
the quark masses are obtained
from Eqs.\ (\ref{mqm0mI}) and (\ref{mIfin}):
\begin{equation} \label{mqfin}
m_q=m_{q0}+\frac{D}{n_{\mathrm{b}}^z}
  \left[
1-\frac{8T}{\lambda T_c} \exp\left(-\lambda \frac{T_\mathrm{c}}{T}\right)
  \right].
\end{equation}
The corresponding partial derivatives are easy to get
\begin{eqnarray}
\frac{\partial m_q}{\partial n_{\mathrm{b}}}
&=& -\frac{zD}{n_{\mathrm{b}}^{z+1}}
  \left[
1-\frac{8T}{\lambda T_c} \exp\left(-\lambda \frac{T_\mathrm{c}}{T}\right)
  \right],   \\
\frac{\partial m_q}{\partial T}
&=&
 -\frac{8D}{n_{\mathrm{b}}^z}
  \left[
\frac{1}{\lambda T_{\mathrm{c}}}+\frac{1}{T}
  \right] \exp\left(-\lambda\frac{T_{\mathrm{c}}}{T}\right).
    \label{dmqdTexp}
\end{eqnarray}
The chemical potentials $\mu_i$ $(i=u, d, s, e)$ are obtained
by solving Eqs. (\ref{eq1})-(\ref{chg0}), all other
thermodynamic quantities can then be obtained.

Figure \ref{S13nb} shows the entropy per baryon as a function
of temperature for different densities. It is an increasing
function of the temperature and goes to zero at zero temperature.
This is ensured by the fact that we have
$\lim_{T\rightarrow 0}\partial m_q/\partial T=0$
from Eq.\ (\ref{dmqdTexp}). It is interesting to note that
we did not require this in deriving
the scaling Eq.\ (\ref{mqfin}) in Sec.\ \ref{qmass}.
We got this automatically and naturally.

\begin{figure}
\epsfxsize=8.2cm
\epsfbox{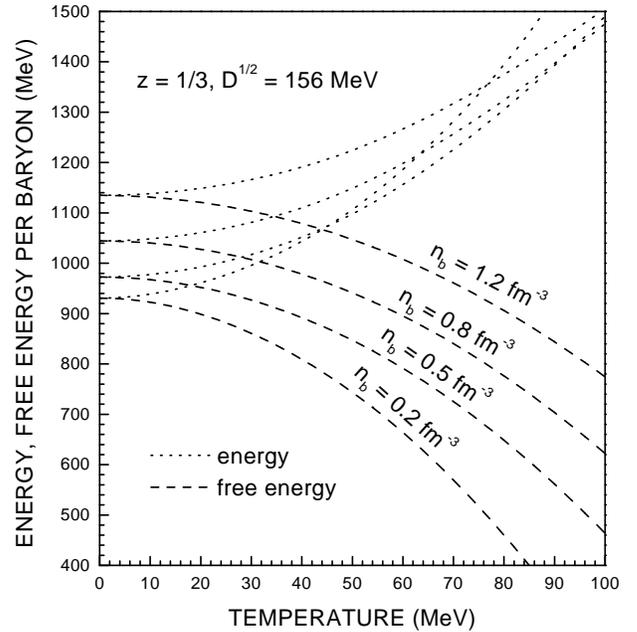}
\caption{
  Temperature dependence of the energy and free energy
  per baryon of SQM. The energy is an increasing function
  of the temperature while the free energy decreases with
  increasing temperature.
         }
\label{enb13t}
\end{figure}

Figure \ref{enb13t} gives the temperature dependence of
the energy and free energy. It is obvious that the energy
is an increasing function of temperature while the free energy
decreases with increasing temperature.

\begin{figure}
\epsfxsize=8.2cm
\epsfbox{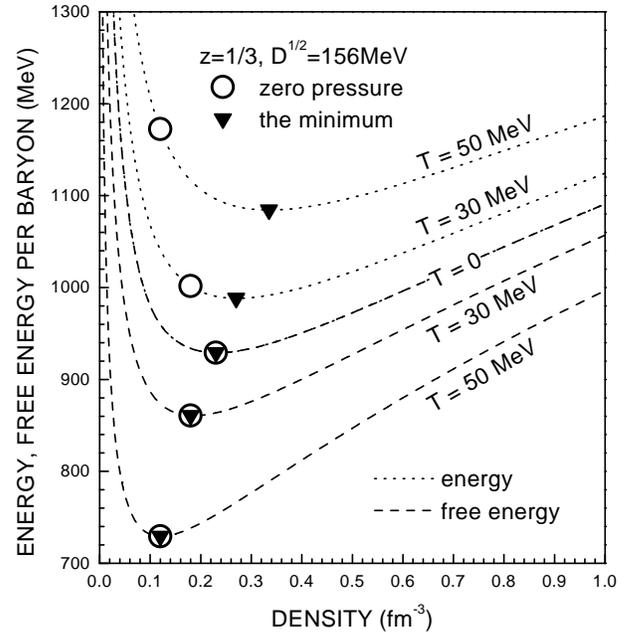}
\caption{
   Density dependence of the energy and free energy
   per baryon of SQM at different temperature.
   On every free energy lines, the zero pressure point
   is exactly located at the minimum. However, they
   are generally not at the same points for the energy
   per baryon, except zero temperature where the energy
  and free energy is identical.
        }
\label{enb13}
\end{figure}

In Fig.\ \ref{enb13}, we plot the density dependence of
the energy and free energy per baryon at different temperature.
The free energy minimum corresponds exactly to the zero pressure,
satisfying Eq.\ (\ref{pnb}). However, these two points (zero pressure
and the minimum) are generally not the same for the energy
per baryon at finite temperature. But at zero temperature
they coincide because the energy and free energy are equal
to each other at zero temperature.

\begin{figure}
\epsfxsize=8.2cm \epsfbox{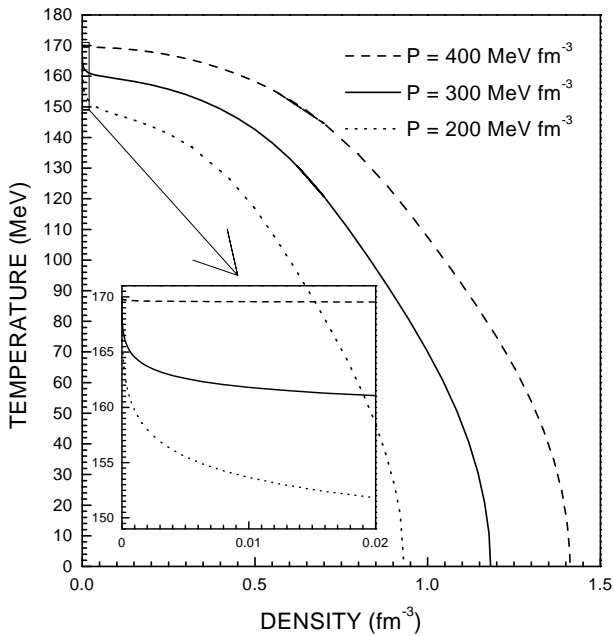} \caption{
  Temperature vs baryon number densities for different pressure values.
  All lines go to the deconfinement temperature $T_{\mathrm{c}}$
  as the density  approaches to zero. But at zero temperature,
  the densities corresponding to different pressure are different
  from each other, lower pressure corresponding to lower density.
         }
\label{t13nb}
\end{figure}

As pointed out in Ref.\ \cite{Zhangy2002PRC65}, a density and
temperature dependent mass model should have the ability
of investigating phase transition. This
is demonstrated in Fig.\ \ref{t13nb}. The figure is plotted
by adjusting, for a given density, the temperature to such a
value that it gives a definite pressure indicted in the legend.
It is obviously shown that all lines go to the deconfinement
temperature $T_{\mathrm{c}}$ as the density  approaches to zero.
But at zero temperature, the density is different for different
pressure. Higher densities correspond to higher pressure.

\section{Properties of strangelets}
\label{slets}

\begin{figure}
\epsfxsize=8.2cm
\epsfbox{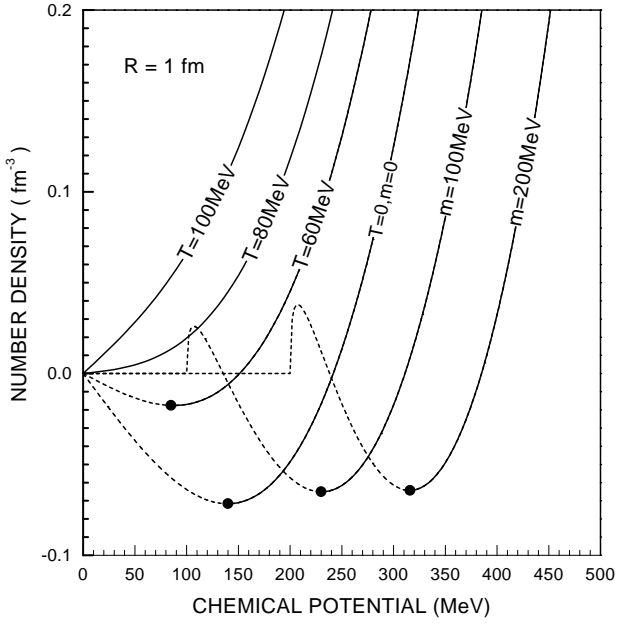}
\caption{
 Chemical potential dependence of the particle number density
at different temperature and mass. The radius is fixed to be
1 fm.
         }
\label{fignmu}
\end{figure}

To study strangelets, the special problem is
to include the finite size effect. We do this by applying
the multi-reflection method, originally comprised
by Balian and Bloch \cite{Balian1970AP60}, later developed
by Madsen \cite{Madsen1993PRL70}, Farhi, Berger, and Jaffe
 \cite{Farhi1984PRD30,Berger1987PRC35} etc.,
and applied to the mass density and temperature dependent model
by Zhang and Su \cite{Zhangy2003MPLA18,Zhangy2003PRC67,Zhangy2004JPG30}.
We express the quasiparticle contribution to
the thermodynamic potential density of strangelets as
$
\Omega=\sum_i\Omega_i(T,\mu_i,m_i,R)
$\ 
with
\begin{eqnarray}
\Omega_i
&=&
  -T\int_0^{\infty}
    \left\{
 \ln\left[1+e^{-(\sqrt{p^2+m_i^2}-\mu_i)/T}\right]
    \right. \nonumber\\
&&  \left. \phantom{-T}
+\ln\left[1+e^{-(\sqrt{p^2+m_i^2}+\mu_i)/T}\right]
    \right\}
n_i^{\prime}\mbox{d}p
\end{eqnarray}
where the density of state $n_i^{\prime}(p,m_i,R)$ is given
in the multi-expansion approach \cite{Balian1970AP60} by
\begin{equation} \label{tmd}
n_i^{\prime}(p,m_i,R)
  =g_i\left[
      \frac{p^2}{2\pi^2}
      +\frac{3p}{R} f_{\mathrm{S}}\left(x_i\right)
      +\frac{6}{R^2} f_{\mathrm{C}}\left(x_i\right)
    \right].
\end{equation}
Here $x_i\equiv m_i/p$, the functions $f_{\mathrm{S}}(x_i)$
 \cite{Farhi1984PRD30,Berger1987PRC35}
and $f_{\mathrm{C}}(x_i)$ \cite{Madsen1993PRL70} are
\begin{equation}
f_{\mathrm{S}}(x_i)=-\frac{1}{4\pi^2}\mbox{arctan}(x_i)
\end{equation}
and
\begin{equation}
f_{\mathrm{C}}(x_i)
=\frac{1}{12\pi^2}\left[1-\frac{3}{2x_i}\mbox{arctan}(x_i)\right].
\end{equation}
Then all other thermodynamic quantities are straightforward
from Eqs.\ (\ref{PMDTD})-(\ref{nMDTD}).
Here is the number density for flavor $i$:
\begin{equation}
n_i=\int_0^{\infty}
     \left[
  \eta_i^+-\eta_i^-
     \right]
    n_i^{\prime}(p,m_i,R)\ \mathrm{d}p,
\end{equation}
where $\eta_i^{\pm}$ is the fermion distribution function, i.e.,
\begin{equation}
\eta_i^{\pm}\equiv
\frac{1}{1+e^{(\sqrt{p^2+m_i^2}\mp\mu_i)/T}}.
\end{equation}
Because we treat the particles and anti-particle as a whole,
the number densities $n_i$ can be both positive and negative
theoretically. A negative particle number means anti-particles.
Fig.\ \ref{fignmu} shows the chemical potential dependence
of the number density for various temperature and mass.
In general, the thermodynamic potential density is
an even function while the number density is an odd function
of the chemical potential, and the number density is zero
at zero chemical potential. If finite size effects can be ignored,
as in Eq.\ (\ref{ni56}) of the preceding section,
this function is monotonically increasing, and so positive
chemical potentials correspond to positive number density.
When the finite size effects are included (the surface term
 \cite{Farhi1984PRD30,Berger1987PRC35}
and curvature term \cite{Madsen1993PRL70}), the case becomes
more complex. When the temperature is very high, the function
is still monotonic. However, when the temperature becomes
lower than some special value, the function becomes
non-monotonic, and the number density is negative
for some special positive chemical potentials.
For smaller masses, the chemical potentials are smaller.
So light quarks do not fall into this region of chemical potentials
in actual cases. However, when quark masses become bigger,
the corresponding chemical potentials becomes also bigger.
It is therefore possible that quarks with comparatively
bigger mass happen to have positive chemical potential and
negative number density in some special cases.
We will see such special examples a little later.
One may concern that non-monotone of the number density
leads to $\partial n/\partial\mu<0$ which violates the stability
condition. However, our results are always located in the
regime where the derivative is positive. The regime is marked
with full lines in Fig.\ \ref{fignmu}.

The free energy density is
\begin{equation}
F=\sum_i\left(
\Omega_i+\mu_i n_i,
      \right),
\end{equation}
the energy density $E$ is
\begin{eqnarray}
E
&=&
\sum_i\left[
 \int_0^{\infty}
\left(
  \eta_i^++\eta_i^-
\right)
\sqrt{p^2+m_i^2}\
n_i^{\prime}(p,m_i,R)\mathrm{d}p
      \right. \nonumber\\
&&    \left. \phantom{\sum_i[}
  -T\frac{\partial\Omega_i}{\partial m_i}
    \frac{\partial m_i}{\partial T}
        \right],
\end{eqnarray}
the entropy density is
\begin{equation}
S=\sum_i\left(
  -\frac{\partial\Omega_i}{\partial T}
  -T\frac{\partial\Omega_i}{\partial m_i}
          \frac{\partial m_i}{\partial T}
        \right),
\end{equation}
and the pressure $P$ is
\begin{equation}
P=\sum_i \left(
 -\Omega_i + n_{\mathrm{b}}\frac{\partial m_i}{\partial n_{\mathrm{b}}}
 \frac{\partial\Omega_i}{\partial m_i}
 -\frac{R}{3}\frac{\partial\Omega_i}{\partial R}
          \right).
\end{equation}
In the above, the relevant partial derivatives are
\begin{equation}
\frac{\partial\Omega_i}{\partial T}
=\int_0^{\infty}
 \ln\left[
    \frac{(1-\eta_i^+)(1-\eta_i^-)}
        {(1/\eta_i^+-1)^{\eta_i^+}(1/\eta_i^--1)^{\eta_i^-}}
    \right]
 n_i^{\prime} 
  \mathrm{d}p,
\end{equation}
\begin{eqnarray}
\frac{R}{3}\frac{\partial\Omega_i}{\partial R}
&=&
  -g_i\int_0^{\infty}
 \left[
  2\sqrt{p^2+m_i^2}
 +T\ln(\eta_i^+\eta_i^-)
 \right]
 \nonumber\\
&& \phantom{-g_i}
     \times
     \left[
\frac{p}{R}f_{\mathrm{S}}(x_i)+\frac{4}{R^2}f_{\mathrm{C}}(x_i)
     \right]
  \mathrm{d}p,
\end{eqnarray}
and
\begin{eqnarray}
\frac{\partial\Omega_i}{\partial m_i}
&=& \int_0^{\infty}
    \left\{
\frac{(\eta_i^++\eta_i^-)n_i^{\prime}}
     {\sqrt{1+p^2/m_i^2}}
    \right. \nonumber\\
&&  \left. \hspace{-0.5cm}
+\left[
  2\sqrt{p^2+m_i^2}
 +T\ln(\eta_i^+\eta_i^-)
\right]
\frac{\partial n_i^{\prime}}{\partial m_i}
    \right\}
 \mathrm{d}p
\end{eqnarray}
with
\begin{equation}
\frac{\partial n_i^{\prime}}{\partial m_i}
=\frac{3g_i}{4\pi^2 m_i R^2}
 \left[
 \frac{p}{m_i}\mbox{arctan}\left(\frac{m_i}{p}\right)
-\frac{1+Rm_i}{1+m_i^2/p^2}
 \right].
\end{equation}

Quark masses and relevant derivatives are still given
by Eqs.\ (\ref{mqfin})-(\ref{dmqdTexp}).

\begin{figure}
\epsfxsize=8.2cm
\epsfbox{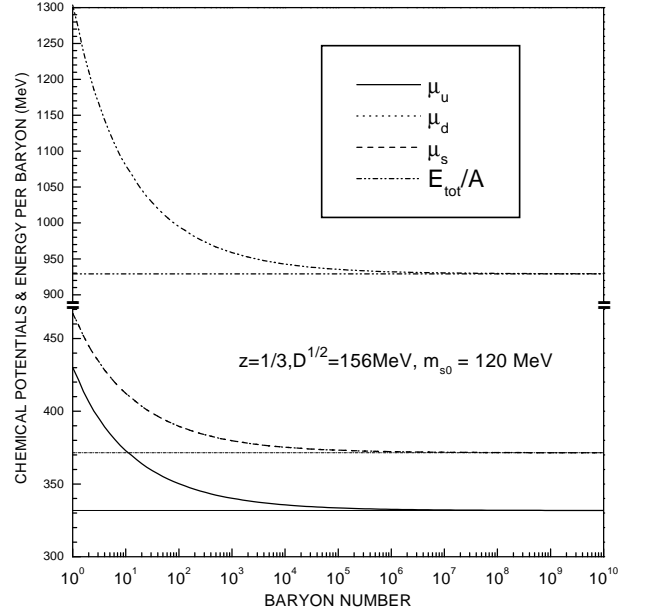}
\caption{
 Chemical potentials and energy per baryon from bulk
 strange quark matter to finite baryon number at
 zero temperature and zero pressure.
         }
\label{emua}
\end{figure}

In Ref.\ \cite{Zhangy2001EPL56}, charge neutrality is also
imposed for strangelets. This is convenient for checking
whether the formulas are continuous from bulk SQM
to finite baryon number. Fig.\ \ref{emua} gives the energy
per baryon and chemical potentials versus baryon number
at zero temperature and zero pressure.
The horizontal lines are the corresponding values for
bulk SQM. It is very clear that all quantities approach
to the corresponding bulk values with increasing baryon number,
and finite size effects destabilize low baryon number strangelets.

Now we treat strangelets in another different way.
Instead of imposing the charge neutrality Eq.\ (\ref{chg0}),
we require that the electron and positron number densities are zero, i.e.,
\begin{equation}
n_e=0
\end{equation}
because the radius of strangelets is much smaller than the Compton
wavelength of electrons and positrons. Electrons and positrons are
not involved in the strong interaction, or in other words, they
are not confined, so the finite size terms in Eq.\ (\ref{tmd})
vanish for them. From Eq.\ (\ref{delnim0}), we have
\begin{equation}
n_e=\frac{\mu_e}{3}\left(T^2+\frac{\mu_e^2}{\pi^2}\right).
\end{equation}
Therefore, zero $n_e$ means zero $\mu_e$.
With a view to the chemical equilibrium
Eqs.\ (\ref{eq1}) and (\ref{eq2}), we naturally get
\begin{equation} \label{muuds}
\mu_u=\mu_d=\mu_s.
\end{equation}
In fact, Eq.\ (\ref{muuds}) is the condition to find out the
configuration, which has the lowest energy per baryon,
from the all possible strangelets with a fixed baryon
number \cite{HeYB53PRC1903}.
Due to Eq.\ (\ref{muuds}), only one chemical potential is
left independent. And it can then be determined by solving
\begin{equation}
\frac{1}{3}(n_u + n_d + n_s)
=\frac{A}{(4/3)\pi R^3}
\end{equation}
for a given baryon number $A$, temperature $T$, and radius $R$.

\begin{figure}
\epsfxsize=8.2cm
\epsfbox{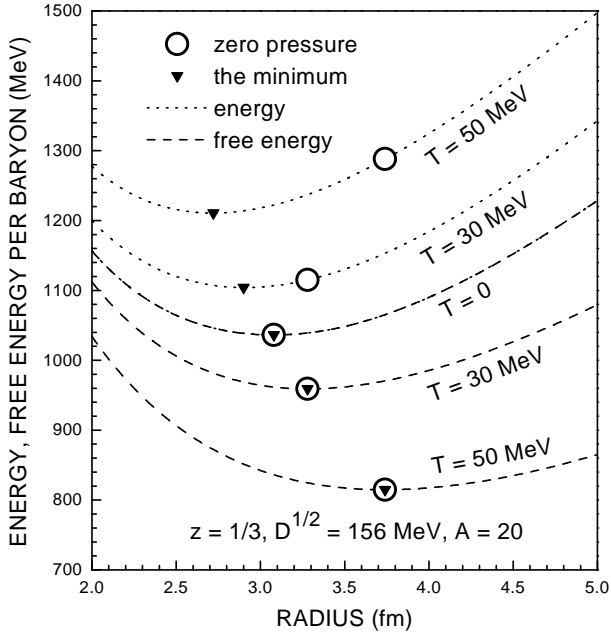}
\caption{
 Energy and free energy vary with the radius at different temperature.
 The mechanically stable radius (where the pressure is zero) is marked
 with a small circle on each line while triangles indicate
 the minimum.
          }
\label{e13r}
\end{figure}

Figure \ref{e13r} shows the energy and free energy
per baryon as a function of the radius for $A=20$ at different temperature.
The points marked with an open circle are the mechanically
stable radius where the pressure is zero. The minimum of each
line is marked with a triangle. Again we see that these two points
are always the same on the free energy line. But they are
different on the energy line at finite temperature. However, they
coincide at zero temperature because the energy and free energy
are equal at zero temperature.

For a given $A$ and $T$, the mechanically stable radius is obtained
by adjusting it so that the free energy is minimized, or,
simply by solving the equation
\begin{equation}
P=0.
\end{equation}

\begin{figure}
\epsfxsize=8.2cm
\epsfbox{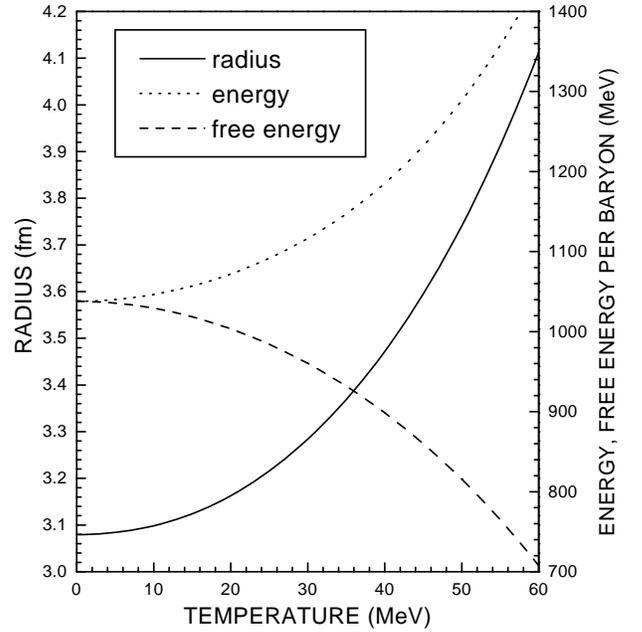}
\caption{
 Temperature dependence of the mechanically stable radius, energy and
 free energy per baryon for the baryon number $A=20$. Both the radius
 and energy per baryon increase with temperature.
         }
\label{r13t20a}
\end{figure}

The temperature dependence of the stable radius for $A=20$
is plotted in Fig.\ \ref{r13t20a} with a solid line. It is obviously
an increasing function of temperature. The corresponding
energy per baryon is also plotted in the same figure, labeled on the
right axis with a dotted line. It is also an increasing function of temperature.
However, the free energy per baryon (dashed line) decreases with increasing
temperature.

\begin{figure}
\epsfxsize=8.2cm
\epsfbox{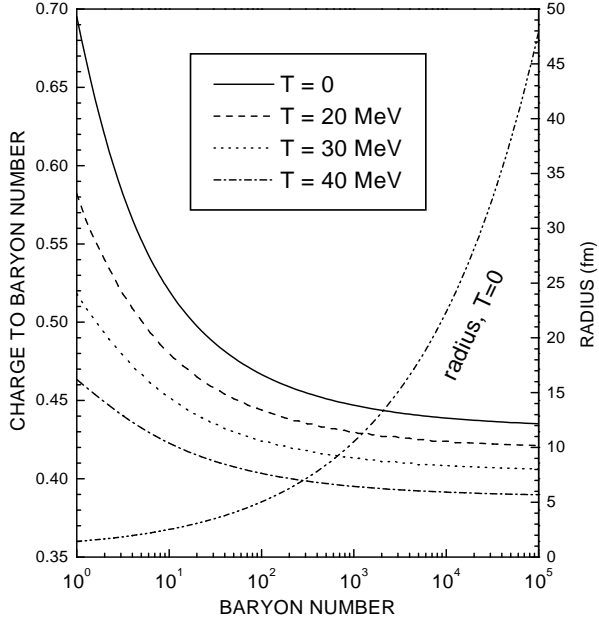}
\caption{
 Charge to baryon number of strangenets at finite temperature.
 At lower baryon numbers, the ratio exceeds 1/2. This means
 a negative number for strange quarks, or the appearance
 of an anti-strange quark. The right axis gives the radius at $T=0$.
        }
\label{r13a}
\end{figure}

Because charge neutrality is not imposed,
strangelets here are charged. The electric charge can be calculated by
\begin{equation}
Z=\frac{4}{3}\pi R^3
 \left(
\frac{2}{3} n_u-\frac{1}{3} n_d-\frac{1}{3} n_s
 \right).
\end{equation}
In Fig.\ \ref{r13a}, we plot the charge to baryon number
ratio as a function of the baryon number at different temperature.
This figure shows that the charge to baryon number ratio
is a decreasing function of the baryon number and temperature.

\begin{figure}
\epsfxsize=8.2cm
\epsfbox{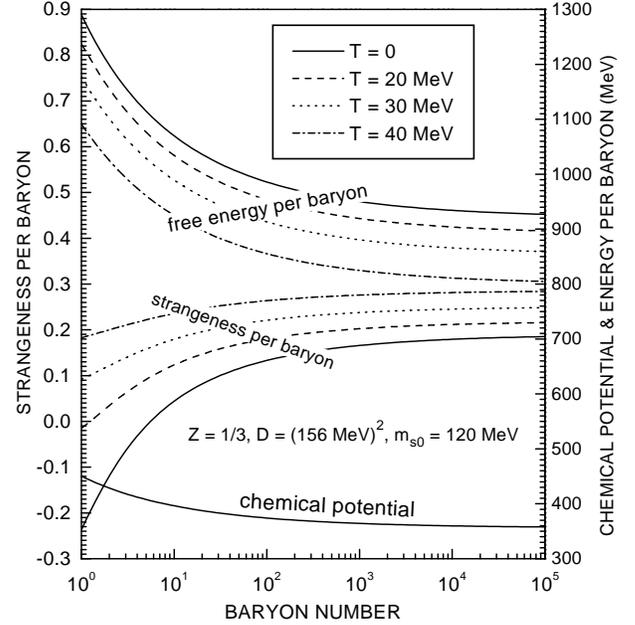}
\caption{
 Strangeness and energy per baryon versus baryon number
 at different temperature. For small baryon numbers at
 very low temperature, the strangeness per baryon becomes negative.
        }
\label{fsa}
\end{figure}

A noticeable feature is that the charge to baryon number ratio
at lower baryon numbers ($A\leq 5$ at zero temperature) becomes
greater than 1/2. This is very different from that in the
bag model where the charge to baryon number ratio is very small.
For normal nuclei, this ratio reaches its biggest value 1/2.
So it seems difficult to understand a heavy positive charge
at first sight.
In fact, it is caused by the fact that $f_s$ is negative
for $1\leq A\leq 5$ and $T=0$, i.e., anti-strange quarks, rather than
strange quarks, appear.
One can see this phenomenon clearly in Fig.\ \ref{fsa} where
the free energy per baryon and the chemical potential have also
been shown. Because of finite size effects, stranelets with
very low baryon numbers are metastable for the parameters chosen.
If one choose a bigger value for the current quark mass,
the value for $D$ should be smaller, and consequently,
the mass for strangelets would be smaller.

\begin{figure}
\epsfxsize=8.2cm
\epsfbox{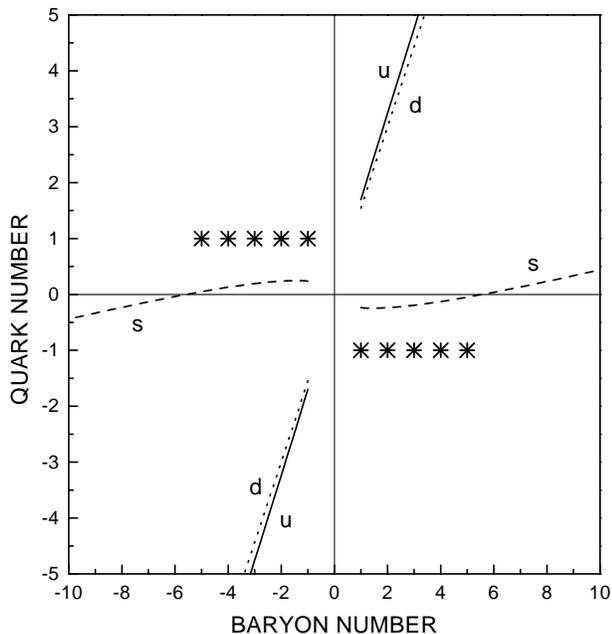}
\caption{
 Quark configuration for low baryon numbers at $T=0$.
 When $1\leq A\leq 5$, the s quark numbers are negative, indicating
 the appearance of anti-strange quarks. The stars are possible
 $\bar{s}$lets in the (A,S) plane.
        }
\label{n13a}
\end{figure}

Fig.\ \ref{n13a} shows the quark configuration for low baryon numbers.
To see the results clearly, we give the corresponding
data in Tab.\ \ref{tabslet}.
The first column is the baryon number,
the second to fourth column are, respectively, the quark numbers
$N_u$, $N_d$, and $N_s$.
 Because shell effects are not taken into account and beta-equilibrium
 is imposed, fractional quark numbers appear in Tab. 1.
Actual quark numbers should naturally be integers. So we approximate
these real numbers to integers by $\mbox{int}(N_i)+N_i/|N_i|$
($i=u, d, s$; int means the number before the decimal point).
The results are shown in the fifth column.
For $A=1$, we have the pentaquark state ($u^2d^2s^{-1}$).
For $A=2$, we have the dibaryon ($u^4d^3s^{-1}$) or octaquark state.
For $A=3$, 4, and 5, we respectively have the multi-quark states
($u^5d^5s^{-1}$), ($u^7d^6s^{-1}$), and ($u^8d^8s^{-1}$). A common feature of these
states is that they all include an anti-strange quark. So we use `$\bar{s}$let'
as the title of the fifth column. The charge number of these $\bar{s}$lets
are given in the sixth column, while the seventh column gives the energies
calculated by the present parameters ($D^{1/2}=156$ MeV and $m_{s0}=120$ MeV)
with perfect $\beta$\ equilibrium.
If one would like to produce 1540 MeV (the actual $\Theta^+$\ resonant mass)
for $A=1$, then one has to take $D^{1/2}=186$ MeV and
get 2856 MeV for $A=2$. So we expect that the mass of
the octaquark ($u^4d^3s^{-1}$), if truely exists, is near 2856 MeV.
For $D^{1/2}=186$ MeV and $A\geq 3$, the strange quark number becomes positive.
So in this case we have only the pentaquark and the octaquark.
Because of uncertainties in parameters,
and also many other factors e.g. the perturbative interaction has not been included
(the quark mass scaling is derived by assuming that the linear confinement
interaction dominates), the concrete values should not be taken seriously,
and further studies are needed.

\begin{table}
\begin{tabular}{|c|c|c|c|c|c|c|c|} \hline
A & $N_u$  &  $N_d$ & $N_s$  & $\bar{\mbox{s}}$let &  Z & $\bar{E}$ (MeV)& $\bar{E}^*$ (MeV)\\ \hline
1 & 1.6951 & 1.5416 & -0.2367 & $u^2d^2s^{-1}$ & 1 & 1290 & 1540 \\ \hline
2 & 3.2390 & 2.9992 & -0.2382 & $u^4d^3s^{-1}$ & 2 & 2397 & 2856 \\ \hline
3 & 4.7570 & 4.4360 & -0.1930 & $u^5d^5s^{-1}$ & 2 & 3471 & 4131 \\ \hline
4 & 6.2634 & 5.8639 & -0.1273 & $u^7d^6s^{-1}$ & 3 & 4528 & 5385 \\ \hline
5 & 7.7624 & 7.2865 & -0.0489 & $u^8d^8s^{-1}$ & 3 & 5573 & 6624 \\ \hline
\end{tabular}
\caption{
 Possible multi-quark states with an anti-strange quark.
 $N_u$, $N_d$, and $N_s$ are the numbers of $u$, $d$, and $s$ quarks,
 satistying the weak equilibrium for the baryon numbers in the
 first column. The fourth column gives the possible $\bar{s}$lets,
 obtained by upgrading the $N_u$, $N_d$, and $N_s$ to integers.
 The sixth column presents the electric charge number,
 while the seventh column is the corresponding masses with
 perfect $\beta$\ equilibrium and $D^{1/2}=156$ MeV.
 The last column is for $D^{1/2}=186$ MeV, in which case
 the s quark number density for $A\geq 3$ becomes positive,
 namely, we have only the first two $\bar{s}$lets:
 the pentaquark and the octaquark.
        }
\label{tabslet}
\end{table}

Recently, the pentaquark state $\Theta^+$(1540)
has aroused a lot of interest \cite{Eidelman2004PLB592}.
The width of $\Theta^+$(1540) is very narrow with upper limit
as small as 9 MeV \cite{Nakano2004MPLA19}. Cahn and trilling have extracted
$\Gamma(\Theta^+)=0.9\pm 0.3$ MeV from an analysis of Xenon bubble chamber
\cite{Cahn2004PRD69}. Although other hadrons like $\phi(1020)$ and
$\Lambda(1520)$ are also narrow [$\Gamma(\phi)=4.26\pm 0.05$ MeV,
$\Gamma(\Lambda)=15.6\pm 1.0$ MeV], we know what makes them narrow.
However, we do not know, until now, why the $\Theta^+$ should be so stable.
We see from the above data that ($u^2d^2s^{-1}$) is the multi-quark state which
mostly satisfy the weak equilibrium for baryon number 1.
This might serve as an explanation for the stability of $\Theta^+$(1540).
Other $\bar{s}$lets, e.g. the dibaryon ($u^4d^3s^{-1}$), an octaquark state,
should also exist, yet to be searched for by experiments though.

In Ref.\ \cite{SchaffnerBielich1997PRD55},
strangelets were also calculated to be heavily charged.
But the electric charge is negative.
There the investigation was concerned with how small metastable
strangelets look like and might decay for different lifetime.
With the similar ideas, Ref.\ \cite{Zhangy2003PRC67} studied
strangelets within density-and-temperature dependent quark masses
and extending the findings in Ref.\ \cite{SchaffnerBielich1997PRD55}
to finite temperature. Present investigation concentrate on
deriving thermodynamic formulas and quark mass scaling,
and finding the lowest-energy configuration from the strangelets with
a fixed baryon number. Naturally, the observation of heavily positively charged strangelets,
or multi-quark states with an anti-strange quark,
depends on the value of the parameter $D$. If we took a much larger
$D$ value, the charge to baryon number ratio would also be small,
or the anti-strange quark would not appear. However, bulk SQM
has no chance to be absolutely stable in that case.

\section{Summary}
\label{sum}

When masses are density and/or temperature dependent,
the thermodynamical formulas are different from that
for constant masses. We have derived a new set of thermodynamical
formulas which can be used to calculate the properties of quark matter
within a density and/or temperature dependent quark mass model.
The new formulas are also instructive when one introduces
a density and/or temperature dependent bag constant in the bag model etc.

We have also argued for a new quark mass scaling at finite temperature.
The basic feature is that quark masses and their partial derivative with
respect to the temperature go to zero when the temperature
approaches to zero. This ensures that all quantities restore
to the density dependent model at zero temperature. It is especially
important that the entropy goes naturally to zero when the temperature
approaches to zero, satisfying the third law of thermodynamics.

With the new thermodynamical formulas and new quark mass scaling,
we have studied the properties of bulk SQM and strangelets.
It is shown that the free energy minimum corresponds exactly to
the zero pressure, both at zero and finite temperature.
The mechanically stable strangelet radius increases with temperature.
An interesting new observation is that low mass strangelets
are heavily positively charged, or appear as multi-quark states
with an anti-strange quark, such as the pentaquark $(u^2d^2\bar{s})$
and the octaquark ($u^4d^3\bar{s}$) etc.,
if bulk SQM is absolutely stable.

\acknowledgments

The authors would like to thank support from the
DOE (DF-FC02-94ER40818) and
NSFC (10375074, 90203004, 10475089, 10435080, 10275037).
G.X.P also acknowledges hospitality at MIT-CTP.
In particular, he is grateful to
Prof.\ E.\ Farhi, R.\ Jackiw, R.\ L.\ Jaffe, J.\ W.\ Negele,
       K.\ Rajagopal, and F.\ Wilczek,
for helpful conversations.

\end{document}